\documentclass[12pt]{article}
\usepackage{amssymb,epsfig}

\begin{document}

\noindent

\def\med{{1\ov 2}}
\def\hepth#1{ {\tt hep-th/#1}}
\def\res{{\rm res}}
\def\min{{\rm min}}
\def\max{{\rm max}}
\def\inv{^{-1}}
\def\ie{{\it i.e. }}%--- i.e.

\def\bd{\begin{description}}
\def\ed{\end{description}}
\def\etal{{\it et al. }}%--- et al.
\def\ie{{\it i.e. }}%--- i.e.
\def\eg{{\it e.g. }}%--- e.g.
\def\Cf{{\it Cf.\ }}%--- Cf.
\def\cf{{\it cf.\ }}%--- cf.
\def\be{\begin{equation}}
\def\ee{\end{equation}}
\def\bes{\begin{equation*}}
\def\ees{\end{equation*}}
\def\beqa{\begin{eqnarray}}
\def\beqas{\begin{eqnarray*}}
\def\eeqa{\end{eqnarray}}
\def\eeqas{\end{eqnarray*}}
\def\bea{\begin{eqnarray}}
\def\eea{\end{eqnarray}}

\newcommand{\comps}{\mathbb{C}}
\newcommand{\reals}{\mathbb{R}}
\newcommand{\integs}{\mathbb{Z}}

\def\om{\omega}
\def\H{{\cal H}}
\def\tr{{\rm tr}}
\def\Tr{{\rm Tr}}
\def\F{{\cal F}}
\def\N{{\cal N}}

\def\d{\partial}
\def\ov{\over}

\def\pder#1#2{\frac{\partial #1}{\partial #2}}%--- partial derivative
\def\der#1#2{\frac{d #1}{d #2}}%--- full derivative
\def\ppder#1#2#3{\frac{\partial^2 #1}{\partial #2\partial #3}}
\def\dpder#1#2{\frac{\partial^2 #1}{\partial #2 ^2 }}
\def\bemat{\left(\begin{array}}
\def\enmat{\end{array}\right)}
\def\theequation{\thesection.\arabic{equation}}
\def\e{\epsilon}
\def\cdotsh{\!\cdot}
\def\cdotsk{\!\cdot\!}

\def\Fpk{\alpha\!\cdot\!\!\F'_k}
\def\Fpu{\alpha\!\cdot\!\!\F'_1}
\def\Fpb{\beta\!\cdot\!\!\F'_1}
\def\Fpd{\alpha\!\cdot\!\!\F'_2}
\def\Fppk{\alpha\!\cdot\!\!\F''_k\!\!\cdot\! \alpha}
\def\Fppu{\alpha\!\cdot\!\!\F''_1\!\!\cdot\! \alpha}
\def\Fppb{\beta\!\cdot\!\!\F''_1\!\!\cdot\! \beta}
\def\Fppd{\alpha\!\cdot\!\!\F''_2\!\!\cdot\! \alpha}
\def\cdotsh{\!\cdot}

%%%%%%%%%%%%%%%%%%%%%%%%%%%%%%%%%%%%%%%%%%%%%%%%%%%%%%%%%%%%%%%%%%%%%%

\begin{titlepage}
\begin{flushright}
{ ~}\vskip -1in
RUNHETC-2000-30 \\
hep-th/0008100\\
August 2000\\
\end{flushright}

\vspace*{20pt}
\bigskip

\centerline{\LARGE Topological Quantum Field Theory }
\bigskip
\centerline{\LARGE and Four-Manifolds\footnote{Talk given at 
the 3rd ECM, Barcelona, July 2000.}}
\vskip 0.9truecm
\centerline{\large\sc Marcos Mari\~no\footnote{marcosm@physics.rutgers.edu} }
\vspace{1pc}

\begin{center}
{\em
New High Energy Theory Center, Rutgers University\\
Piscataway, NJ 08855, USA.} \\

\vspace{5pc}

%%%%%%%%%%%%%%%%%%%%%%%%%%%%%%%%%%%%%%%%%%%%%%%%%%%%%%%%%%%%%%%%%%%%%%%%%%%%%%%
%%%                                Abstract                                 %%%
%%%%%%%%%%%%%%%%%%%%%%%%%%%%%%%%%%%%%%%%%%%%%%%%%%%%%%%%%%%%%%%%%%%%%%%%%%%%%%%
{\large \bf Abstract}

\end{center}

I review some recent results on four-manifold invariants 
which have been obtained in the context of topological 
quantum field theory. I focus on three different aspects: 
(a) the computation of correlation functions, which give 
explicit results for the Donaldson invariants of non-simply 
connected manifolds, and for generalizations of these invariants 
to the gauge group $SU(N)$; (b) compactifications to lower dimensions, and  
relations with three-manifold topology and with intersection theory 
on the moduli space of flat connections on Riemann surfaces; (c) 
four-dimensional 
theories with critical behavior, which give some remarkable 
constraints on Seiberg-Witten invariants and new results on the 
geography of four-manifolds.

\end{titlepage}

%%%%%%%%%%%%%%%%%%%%%%%%%%%%%%%%%%%%%%%%%%%%%%%%%%%%%%%%%%%%%%%%%%%%%%%%%%%%%%%
%%%                                Section 1                                %%%
%%%%%%%%%%%%%%%%%%%%%%%%%%%%%%%%%%%%%%%%%%%%%%%%%%%%%%%%%%%%%%%%%%%%%%%%%%%%%%%

\section{Introduction}

One of the original motivations of Witten \cite{tqft} to introduce
 topological quantum field theories (TQFT) was precisely to understand 
the Donaldson invariants of four-manifolds from a physical point of view. 
This approach proved its full power in 1994, when it was realized that 
all the information of Donaldson theory was contained in the 
 Seiberg-Witten (SW)  invariants. These new invariants 
led to a true revolution 
in four-dimensional topology, and they were introduced in \cite{mfm} 
based on nonperturbative results in supersymmetric quantum field 
theory. The relation between Donaldson invariants and SW 
invariants was fully clarified in an important paper by G. Moore 
and E. Witten \cite{mw}, where they introduce the so called $u$-plane integral. 

In this note, I review some recent results on four-manifold invariants 
which have been obtained through the use of $u$-plane integral techniques. 
I emphasize how these results are related to the physics 
of four-dimensional quantum field theories. First, I discuss Donaldson 
invariants as correlation functions in TQFT. I present some new results 
for non simply connected manifolds (for product ruled 
surfaces, in particular), and for extensions of Donaldson theory 
to higher rank gauge groups. Second, I use compactifications of the 
field theory to make contact with results in three and two dimensions. In the 
two-dimensional case, I recover in fact Thaddeus' celebrated formula for 
the intersection pairings on the moduli 
space of flat connections on a Riemann surface. 
Finally, I consider qualitatively new physics (a field theory 
with critical behavior) to obtain new relations between the SW 
invariants and classical invariants of four-manifolds. In the last 
section, I briefly consider some open problems.

\medskip

\section{Correlation functions}
\setcounter{equation}{0}
\subsection{General aspects}
The Donaldson invariants of smooth, compact, oriented four-manifolds
$X$ \cite{DoKro} are defined by using
intersection theory on the moduli space of anti-self-dual connections. The
cohomology classes
on this space are associated to homology classes of $X$ through the slant
product \cite{DoKro} or, in the
context of topological field theory, by using
the descent procedure \cite{tqft}. Here we will restrict ourselves to
the Donaldson invariants
associated to zero, one and two-homology classes\footnote{The inclusion of
three-classes has been
considered in \cite{mmtwo}.}. Define
\begin{equation}
{\bf A}(X)={\rm Sym}(H_0(X) \oplus H_2(X))\otimes \wedge ^* H_1(X).
\label{ax}
\end{equation}
Then, the Donaldson invariants can be regarded as functionals
\begin{equation}
D^{w_2(E)}_X: {\bf A}(X) \rightarrow {\bf Q},
\label{poly}
\end{equation}
where $w_2(E) \in H^2(X, {\bf Z})$ is the second Stiefel-Whitney class
of the gauge bundle. It is convenient to organize these invariants as follows.
Let $\{\delta_i\}_{i=1,\ldots,b_1}$ be a basis of one-cycles,
$\{\beta_i\}_{i=1,\ldots,b_1}$ the corresponding dual basis of
harmonic one-forms, and
$\{S_i\}_{i=1,\ldots, b_2}$ a basis of two-cycles. We introduce
the formal sums
\begin{equation}
\delta= \sum_{i=1}^{b_1} \zeta_i \,  \delta_i , \qquad\qquad S=
\sum_{j=1}^{b_2}
v_i \,S_i,
\label{cycles}
\end{equation}
where  $v_i$ are complex
numbers, and $\zeta_i$ are Grassmann variables. The generator of the
$0$-class will
be denoted by $x \in H_0(X, {\bf Z})$. We then define the Donaldson-Witten
generating function:
\begin{equation}
Z_{DW}(p, \zeta_i,v_i)=D_X^{w_2(E)}({\rm e}^{p x + \delta + S}),
\label{donwi}
\end{equation}
so that the Donaldson invariants can be read off from the expansion of the
left-hand side in powers of $p$, $\zeta_i$ and $v_i$. The main result
 in \cite{tqft}
is that $Z_{DW}$ can be understood as the generating
functional of a twisted version of
the $N=2$ supersymmetric gauge theory -- with gauge group $SU(2)$ -- in
four dimensions.
In the twisted theory one can define observables $O(x)$, $
I_1(\delta)=\int_{\delta} O_{1}$,
$I_2(S)=\int_{S}O_{2}$ (where
$O_i$ are functionals  of the fields of the theory) in one to
one correspondence with the homology classes of $X$, and in such a
way that the generating functional
$$
\langle {\rm e}^{pO(x)+ I_1(\delta)+ I_2(S)}\rangle
$$
is precisely $Z_{DW}(p,\zeta_i,v_i)$.

Based on the low-energy effective
descriptions of $N=2$ gauge theories obtained in \cite{sw}, Witten
obtained
a explicit formula for (\ref{donwi}) in terms of SW invariants for
manifolds
of $b_2^+>1$ and simple type \cite{mfm}. The general framework 
to give a complete
evaluation
of (\ref{donwi}) was established in \cite{mw}. The main result of 
Moore and Witten is
an explicit
expression for the generating function
$Z_{DW}$:
\begin{equation}
Z_{DW}=Z_u+Z_{SW}
\label{donwii}
\end{equation}
which consists of two pieces. $Z_{SW}$ is the contribution from the moduli
space $M_{SW}$ of solutions of the SW monopole equations.
$Z_u$ (the $u$-plane integral henceforth) is the integral of a
certain modular form over the fundamental domain of the group $\Gamma^{0}(4)$,
that is, over the quotient $\Gamma^{0}(4)\setminus H$,
where $H$ is the upper half-plane. The explicit form of $Z_u$ was
derived
in \cite{mw} for simply connected four-manifolds, and extended to the
non-simply connected case in \cite{mmtwo}. $Z_u$ is non-vanishing only for
manifolds with $b_2^{+}=1$, and provides a simple physical explanation of the
failure of topological invariance of the Donaldson invariants on those
manifolds \cite{mw}.

\subsection{Donaldson invariants in the non-simply connected case}

Most of the computations of Donaldson invariants have focused on 
simply connected manifolds. The study of the nonsimply connected side 
was initiated in \cite{mw,lns}, and finally a complete description of the 
invariants was given in \cite{mmtwo}. Some additional results were obtained 
in \cite{lozm}. The nonsimply connected case presents some new features,
 mostly when 
$b_2^+=1$.  
Of particular interest are the Donaldson invariants of product 
ruled surfaces ${\bf S}^2 \times \Sigma_g$, which as far as I know have not 
been completely determined from a mathematical point of view.
 Recall that the invariants 
depend on the 
chamber chosen in the K\"ahler cone. The result gets simpler in the 
limiting chambers of very small 
or large volumes for ${\bf S}^2$. We will take a symplectic basis 
of one cycles in $\Sigma_g$, $\delta_i$, $i=1, \cdots, 2g$, and consider 
the ${\rm Sp}(2g, {\bf Z})$-invariant element 
$\iota=-2\sum_{i=1}^g \delta_i \delta_{i+g}$. In the limit of small 
volume for 
${\bf S}^2$, the generating functions $Z_g^{w_2(E)}=
D^{w_2(E)}_{{\bf S}^2 \times \Sigma_g}({\rm e}^{px + r
\iota+ s\Sigma_g + t{\bf S}^2})$ are given by \cite{mmtwo,lozm}:
\begin{eqnarray}
&Z^{w_2(E)=0}_{g}=-{i\over 4}
\biggl[ (h_{\infty}^2 f_{2\infty})^{-1} {\rm e} ^{2pu_{\infty} +
2st T_{\infty}}
\Bigl( 2 f_{1\infty} h_{\infty}^2 s+2r\Bigr)^g
\coth \Bigl( {i s \over 2h_{\infty}} \Bigr)
\biggr]_{q^0}, \label{ruledexp}\\
&Z^{w_2(E)=[{\bf S}^2]}_{g, {\bf S}^2}=-{1\over 4}
\biggl[ (h_{\infty}^2 f_{2\infty})^{-1} {\rm e} ^{2pu_{\infty} +
2st T_{\infty}}
\Bigl( 2 f_{1\infty} h_{\infty}^2 s+2r\Bigr)^g
\csc \Bigl( { s \over 2h_{\infty}} \Bigr)
\biggr]_{q^0},
\label{ruledexpf}
\end{eqnarray}
and they vanish for the other choices of Stiefel-Whitney class.
For $g=0$, one recovers the expressions
for ${\bf S}^2 \times {\bf S}^2$ which were obtained in \cite{mw,gottzag}. 
The 
above equations involve the modular forms with $q$ -expansions:
\begin{eqnarray}
&  u_{\infty}= {1\over 2} {{\vartheta_2^4+\vartheta_3^4}\over
(\vartheta_2\vartheta_3)^2}={1\over 8 q^{1/4}}(1+20
q^{1/2}-62q+\cdots),\nonumber\\
& T_{\infty} = -{1\over24}\left(  {E_2\over
h^2_{\infty}}-8u_{\infty}\right)= q^{1/4}(1-2 q^{1/2}+6q+\cdots)
,\nonumber\\
& h_{\infty}(\tau)= {1
\over 2}
\vartheta_2
\vartheta_3=q^{1/8}(1+2q^{1/2}+q+\cdots),\nonumber\\
& f_{1\infty}(q)= {2 E_2+
\vartheta_2^4 +
\vartheta_3^4 \over 3
\vartheta_4^8}= 1+ 24 q^{1/2} + \cdots,\nonumber\\ 
& f_{2\infty}(q)= {\vartheta_2
\vartheta_3 \over  2
\vartheta_4^8}= q^{1/8} + 18 q^{5/8}+
\cdots,
\label{modforms}
\end{eqnarray}
and the subscript $q^0$ means that one has to extract the 
$q^0$ power in the $q$-expansion. The expressions for the other 
limiting chamber can be found in two ways: since 
the wall-crossing formula in the nonsimply connected case 
was obtained in \cite{mmtwo}, one can sum to the above 
expression an infinite number of wall-crossing terms. 
Alternatively, one can perform a direct evaluation of 
the $u$-plane integral \cite{lozm}. The Donaldson invariants 
for the chamber of small volume for $\Sigma_g$ can also be computed using 
the structure of the Floer cohomology of $\Sigma_g \times {\bf S}^1$ 
\cite{vm}, and results are in full agreement with the 
generating functions obtained in the 
context of the $u$-plane. 

\subsection{Extension to gauge group $SU(N)$}

The Donaldson invariants are usually defined for the gauge group $SU(2)$. 
In principle, one can formally consider invariants of four-manifolds 
defined from anti-self dual $SU(N)$ gauge connections. 
Although this seems to be 
pretty difficult from a mathematical point of view, the evaluation of 
the would-be $SU(N)$ Donaldson invariants turns out to be tractable 
using quantum field theory \cite{mmone}. The result is simpler for manifolds 
of simple type and with $b_2^+>1$. Not surprisingly, it can be 
expressed in terms of the cohomology ring of $X$ and of SW  
invariants: 
\begin{eqnarray}
&\langle {\rm e} ^{pO(x) + I_2(S)} \rangle_{SU(N)} = 
{ \alpha}_N^\chi
{ \beta}_N^\sigma \sum_{k=0}^{N-1} \omega^{k(N^2-1)
\chi_h } \sum_{\lambda^I} \prod_{I=1}^{N-1}
SW(\lambda^I)\nonumber\\
&  \cdot \Bigl( \prod_{1\le I< J \le N-1}
 q_{IJ}^{-(\lambda^I, \lambda^J)} \Bigr) \exp \biggl[
p\omega^{2k}N + 2\omega^{2k} S^2 + 2 \omega^k \sum_{I=1}^{N-1} (S,
\lambda^I)  \sin {\pi I \over N} \biggr],
\label{simple}
\end{eqnarray}
where $\omega=\exp[i \pi /N]$, $\chi_h=(\chi + \sigma)/ 4$, and 
$\chi$, $\sigma$ are the Euler characteristic and the signature of 
$X$, respectively. The $q_{IJ}$ are 
$\exp{\pi i \tau_{IJ}}$, where $\tau_{IJ}$, $I\not=J$, are 
the leading terms of the 
offdiagonal effective couplings $\tau_{IJ}$, which have been computed in 
\cite{em}. The sum in (\ref{simple}) is over basic classes, and 
$(\,,\,)$ is the intersection form of $X$. Finally, 
$\alpha_N$ and $ \beta_N$ are universal constants. 
In the above expression we have only considered $SU(N)$ bundles with 
zero Stiefel-Whitney class. In addition, one can consider additional 
operators associated to higher Casimirs of the gauge group, that we have not 
included in (\ref{simple}). Notice that the above expression shows that 
the theory factorizes down to the ``magnetic'' 
Cartan torus $U(1)^{N-1}$, but there is 
an important mixing measured by the off-diagonal effective couplings.

\section{Compactification}

\subsection{Down to three dimensions}

To make contact with results in three-dimensional topology, one 
should consider four-manifolds of the form $X={\bf S}^1\times Y$. 
Donaldson theory on these manifolds has been explored in 
\cite{mmthree}. Using results from supersymmetric gauge theory, 
we would expect the partition function of 
Donaldson-Witten theory on $Y\times {\bf S}^1$ for gauge group $G$ 
to agree with the 
Rozansky-Witten invariant $Z_{RW}(Y,X_{G})$ \cite{rw}, where $X_G$ is 
a hyperK\"ahler manifold. When $G=SU(2)$, $X_{SU(2)}$ is the Atiyah-Hitchin 
manifold and the Rozansky-Witten invariant is the Casson-Walker-Lescop 
invariant $\lambda_{\rm CWL}(Y)$. For $G=SU(N)$, $X_{SU(N)}$ is 
the reduced moduli space of $N$ monopoles, which is a hyperK\"ahler manifold 
of dimension $4(N-1)$. 

This expectation can be partially checked. Using (\ref{simple}) 
and the Meng-Taubes theorem \cite{mt}, 
one can prove that, for $b_1(Y)>1$
\begin{equation}
Z^{SU(N)}_{DW}(Y \times {\bf S}^1)=N^2(\lambda_{\rm CWL}(Y))^{N-1},
\label{three}
\end{equation}
and the left hand side is in fact (up to an overall constant) 
$Z_{RW}(Y,X_{SU(N)})$, which has been recently computed by Habegger 
and Thompson \cite{ht}. 
This gives an interesting non-trivial check of (\ref{simple}). 
For $b_1(Y)=1$ there are 
important subtleties in the correspondence with Rozansky-Witten theory, 
which have been discussed in \cite{mmthree} when the gauge group is $SU(2)$. 

\subsection{Down to two dimensions}
The connection to two-dimensional moduli problems appears 
when one considers product ruled surfaces $X={\bf S}^2 \times \Sigma_g$. 
Anti-self dual connections on $X={\bf S}^2 \times \Sigma_g$ with zero 
instanton number and $w_2(E)=[{\bf S}_2]$ 
are in one-to-one correspondence with flat connections 
on $\Sigma_g$ with odd degree, which form a moduli space $M_g$. 
Donaldson invariants correspond to 
intersection pairings on $M_g$, which were 
determined by Thaddeus in \cite{th}.  
The ${\rm Sp}(2g, {\bf Z})$-invariant cohomology ring of $M_g$ is 
generated by cohomology classes $\alpha$, $\beta$ and $\gamma$, 
of degrees $2$, $4$ 
and $6$, respectively. The relation between the intersection pairings and 
the Donaldson invariants of product ruled surfaces is given by:  
\begin{equation}
\langle \alpha^m \beta^n \gamma^p \rangle_{M_g}=
-D_{{\bf S}^2 \times \Sigma_g}^{w_2(E)=[{\bf S}^2]}
((2\Sigma_g)^m (-4x)^n \iota^p),
\label{pairings}
\end{equation}
where the overall minus sign is due to a different choice of orientation. 
On the other hand, we know the explicit expression for the Donaldson 
invariants, which is given in (\ref{ruledexpf}), and we can then 
rederive some important results about the intersection pairings
 \cite{lozm}.  
The first thing that we can prove is the
recursive relation for insertions of $\gamma$. One easily sees that
\begin{equation}
{\partial \over \partial r}Z^{w_2(E)=[{\bf S}^2]}_{g}=2g
Z^{w_2(E)=[{\bf S}^2]}_{g-1},
\label{rec}
\end{equation}
and this implies, using (\ref{pairings}), that 
$\langle \alpha^m \beta^n \gamma^p \rangle_{M_g}=2g \langle \alpha^m
\beta^n \gamma^{p-1} \rangle_{M_{g-1}}$, 
which is precisely Thaddeus' recursive relation.

We now compute the intersection pairings $\langle \alpha^m \beta ^n \rangle$.
To do this, we use the expansion:
\begin{equation}
\csc z = \sum_{k=0}^{\infty} (-1)^{k+1}(2^{2k}-2) B_{2k}{z^{2k-1} \over
(2k)!},
\label{expacsc}
\end{equation}
where $B_{2k}$ are the Bernoulli numbers. We have to extract now the powers
$s^m p^n$ from the generating function (\ref{ruledexpf}). Fortunately, only 
the leading terms contribute in the $q$-expansion of the modular 
forms. Taking into account
the comparison factors from (\ref{pairings}), and the dimensional constraint
$2m+4n=6g-6$, one finds
\begin{equation}
\langle \alpha^m \beta^n \rangle=(-1)^g { m! \over (m-g+1)!}2^{2g-2}
(2^{m-g+1}-2)  B_{m-g+1},
\label{michael}
\end{equation}
which is exactly Thaddeus' formula for the intersection pairings.

The relation between topological Yang-Mills theory on ${\bf S}^2 \times 
\Sigma_g$ and two-dimensional moduli problems is in fact more interesting, 
since the Donaldson invariants in the limiting chamber of small volume 
for $\Sigma_g$ correspond to the Gromov-Witten invariants of $M_g$. 
We refer the reader to \cite{lozm,vm} for results in this direction.

\section{Critical behavior}

\subsection{Superconformal points}

When one considers topological quantum field theories in four dimensions 
with qualitative new physics, one also finds a completely different kind 
of predictions from a mathematical point of view. In \cite{mmp} we studied 
a quantum field theory with a critical behavior on a four-manifold $X$ 
of simple type and with $b_2^+>1$, namely 
twisted $N=2$ supersymmetric QCD with gauge group $SU(2)$ and 
one massive hypermultiplet with mass 
$m$. It is known 
\cite{apsw} that the low-energy theory becomes superconformal for a certain 
critical value of the mass $m_*$, and that the quantities that 
characterize the theory (like the masses of the BPS 
particles) have a scaling behavior near the critical point. The theory 
has the same BRST operators than topological Yang-Mills theory, although 
mathematically it describes equivariant intersection theory on the moduli 
space of $SU(2)$ monopoles (see \cite{tesis} for a review). Using the 
results of \cite{mw} and some additional input, one can compute the analog 
of the generating function (\ref{donwi}) for this theory, which now depends 
on the extra parameter $m$. To write the result, we need the family of 
Seiberg-Witten elliptic curves for the $N_f=1$ theory \cite{sw}, 
parameterized by $(u,m)\in {\bf C}^2$ and given by:
\begin{equation}
y^2 = x^2(x-u) + 2m x -1 .
\label{curve}
\end{equation}
The curve is easily put into standard form
$y^2 = 4x^3 - g_2 x - g_3$, with
$g_2(u;m) = {4 \over  3} (u^2-6m)$,
$g_3(u;m) = {1 \over  27}(8 u^3 -72 m u + 108)$, and
discriminant $\Delta(u;m) = g_2^3 - 27 g_3^2$.
This discriminant is a cubic in $u$ and has
three roots $u_j(m)$, $j=1,2,3$. For generic,
but fixed, values of $m$
one of the periods of (\ref{curve}) goes to infinity
as $u \rightarrow u_j$ while the other period,
$\varpi_j\equiv \varpi(u_j(m);m)$ remains finite, and in fact is given
by $(\varpi_j )^2 = g_2/(36 g_3)$. The generating function of the 
critical theory
is given by a sum over the
singular fibers of the Weierstrass family
(\ref{curve}) and over the basic classes of $X$:
\begin{eqnarray}
&Z(p,S;m) = k \sum_{j=1}^3 \biggl({g_2^3(u_j(m);m) \over
\Delta'(u_j(m);m) } \biggr)^{\chi_h} (\varpi_j(m))^{7 \chi_h - c_1^2}
\nonumber\\
& \cdot \sum_x  SW(x) (-1)^{(\upsilon^2 + \upsilon \cdot x)/2}
\exp\biggl[ 2 p u_j + S^2 T_j - i {  (S,x) \over  2\varpi_j  }  \biggr] 
\label{genfun}
\end{eqnarray}
Here $\Delta' = { \partial \over  \partial u} \Delta$,
$T_j=-{1 \over 24} \bigl(
(\varpi_j)^{-2}  -8 u_j \bigr) $,
and $k$ is a nonvanishing constant, independent of
$p,S,m$. The topological data of the manifold $X$ enter through $\upsilon$, 
which is an integral lifting of $w_2(X)$, the basic classes $x$ and their 
SW invariants, and the numerical invariants $\chi_h$ and 
$c_1^2=2\chi + 3\sigma$.

The critical behavior of this theory is associated 
to the cusp singularity of (\ref{curve}) when
$m_{*}= {3 \over  2}$, $u_*= 3$. 
Indeed, when  $z=m-m_* \rightarrow 0$, two of the roots of
$\Delta(u;m)=0$, call them $u_\pm(m)$, coincide, 
and the period $\varpi_\pm $ diverges as $z^{-1/4}$, while
$g_2(u_\pm(m);m) \sim z$  and $\Delta'(u_\pm(m);m)
\sim \delta u_{\pm} \sim z^{3/2}$. At the third
singularity all the various factors in (\ref{genfun}) are
given by nonvanishing analytic series in $z$, but,
evidentally, the contributions from $u_\pm(m)$ contain
factors which are diverging or vanishing as $z \rightarrow 0$.
What can we say about the behavior of the
complete function $Z(p,S,m)$ as $z \rightarrow 0$? 
For physical reasons, we do not expect any divergence 
in the correlation functions: there are no infrared divergences 
in spacetime, since $X$ is compact, and since the moduli 
space of vacua is also compact for $b_2^+>1$, we do not expect 
any divergence from the target geometry. In more physical terms, 
since we are working at finite volume, correlation functions should 
still be finite near the critical point. This implies that $Z(p,S,m)$ {\it must be a regular analytic function of $z$ near $z=0$}. 

\subsection{Mathematical implications}

Let's now see what are the mathematical implications of this fact. 
We first define the ``twisted'' Seiberg-Witten series as follows. 
\begin{equation}
SW_X^{w_2 (X)} (z) := \sum_{x} (-1)^{\upsilon^2 + \upsilon\cdot x \over 2} SW
(x) {\rm e}^{z  x}.
\label{series}\end{equation}
This is a finite sum \cite{mfm}. Notice that a change of lifting 
changes (\ref{series}) only by a sign. We now make the key definition:

{\bf Definition}. Let $X$ be a compact, oriented 4-manifold of simple type
with
$b_2^+>1$. We say that ``$X$   is  SST'' (superconformal simple type) if
$SW_X^{w_2 (X)} (z)$ has a zero at $z=0$ of order $\ge \chi_h - c_1^2 -3$.

One has the following result, whose proof can be found in \cite{mmp}:

{\bf Theorem}. If $X$ is SST, then $Z(p,S,m)$ is regular at $m=m_*$.

It is interesting to notice that, for most of the SST manifolds, 
the contributions to $Z(p,S,m)$ from the colliding singularities 
$u_{\pm}$ go to infinity separately as $z\rightarrow 0$, but when 
we sum the two contributions (and we are forced to do that 
because the manifold is compact) the infinities cancel and we get a finite 
result.

All the simple type, four-manifolds we are aware of are in fact SST. 
Using the definition, one 
can check that SST manifolds satisfy the following remarkable property, which gives a relation between SW invariants and the problem of 
geography for four-manifolds:

{\bf Theorem} (Generalized Noether inequality). Let $X$ be SST. If $X$ has
$B$ distinct basic
classes and $B>0$, then
$$
B \ge \biggl[ {\chi_h -c_1^2 \over 2} \biggr].$$
In particular, $c_1^2 \ge \chi_h -2B-1$.

Although being SST is only a sufficient condition for $Z(p,S,m)$ to be 
finite, the analysis of \cite{mmp} leads naturally to the following 
conjecture: 

{\bf Conjecture}. All compact four-manifolds of simple type 
and with $b_2^+>1$ are SST. 

In fact, Feehan, Kronheimer, Lenness and Mrowka have proved in \cite{fklm} 
that the above conjecture is true under some mild assumptions, by using 
the $PU(2)$ monopole equations.  

\section{Conclusions and open problems}

I think it is fair to say that we have a rather complete understanding of  
the relation between Donaldson theory and TQFT 
in four dimensions. There are 
however a few open problems that deserve investigation, both in physics 
and in mathematics: 

1) There are many predictions from TQFT that should still be checked 
from the mathematical side, and I think that this is interesting by itself. 
For example, the results (\ref{ruledexp}) and (\ref{ruledexpf}), as 
well as the wall-crossing formula of \cite{mmtwo} for nonsimply 
connected manifolds, may be 
obtained by generalizing \cite{gottzag}. The extension to 
$SU(N)$ seems still out of reach mathematically, but it would be extremely 
interesting to check (\ref{simple}) in some detail. One can invert the logic 
and say that (\ref{simple}) gives a good reason {\it not} to study the 
$SU(N)$ Donaldson invariants, since it shows that these 
generalizations have the same topological information than the 
SW invariants!

2) In a different direction, it would be interesting to study the theory 
for four-manifolds with $b_2^+=0$. 
A motivation to do that would be to shed some four-dimensional light 
(via compactification on a circle) on the relation 
between the Casson invariant and the three-dimensional Seiberg-Witten 
invariant for homology three-spheres.

3) Finally, the twisted counterparts 
of superconformal field theories in four dimensions certainly 
deserve closer scrutiny.


\begin{thebibliography}{99}

\itemsep=\smallskipamount

\bibitem{apsw} P.C. Argyres, M.R. Plesser, N. Seiberg and E. Witten, 
``New $N=2$ superconformal field theories in four dimensions,'' 
hep-th/9511114, Nucl. Phys. {\bf B 461} (1996) 71.  

\bibitem{DoKro} S.K. Donaldson and P.B. Kronheimer, {\it
The geometry of four-manifolds}, Oxford, 1990.


\bibitem{em} J.D. Edelstein and J. Mas, ``Strong-coupling expansion and 
Seiberg-Witten-Whitham equations,'' hep-th/9901006, Phys. Lett. {\bf B 452} 
(1999) 69. J.D. Edelstein, M. G\'omez-Reino and M. Mari\~no, ``Blowup 
formulae in Donaldson-Witten theory and integrable hierarchies,'' 
hep-th/0006113.

\bibitem{fklm} P.M.N. Feehan, P.B. Kronheimer, T.G. Lenness, and 
T. Mrowka, ``$PU(2)$ monopoles and a conjecture of Mari\~no, Moore and 
Peradze,'' math.DG/9812125, Math. Res. Letters {\bf 6} (1999) 169.
 
\bibitem{gottzag} L. G\"ottsche, ``Modular forms and Donaldson
invariants for 4-manifolds with $b_+=1$,'' alg-geom/9506018; J. Am. Math. Soc.
{\bf 9}
(1996) 827. L. G\"ottsche and D. Zagier,
``Jacobi forms and the structure of Donaldson
invariants for 4-manifolds with $b_+=1$,''
alg-geom/9612020, Selecta. Math. (N.S.) {\bf 4} (1998) 69.

\bibitem{ht} N. Habegger and G. Thompson, ``The universal perturbative 
quantum three-manifold invariant, Rozansky-Witten invariants and the 
generalized Casson invariant,'' math.DG/9911049. 


\bibitem{lozm} C. Lozano and M. Mari\~no, ``Donaldson invariants of 
product ruled surfaces and two-dimensional gauge theories,'' hep-th/9907165, 
to appear in Commun. Math. Phys.  



\bibitem{lns} A. Losev, N. Nekrasov, and S. Shatashvili, ``Issues in
topological gauge theory," hep-th/9711108,
Nucl. Phys. {\bf B 549} (1998); ``Testing Seiberg-Witten solution,"
hep-th/9801061.

\bibitem{tesis} M. Mari\~no, ``The geometry of supersymmetric 
gauge theories in four dimensions,'' hep-th/9701128.


\bibitem{mmone} M. Mari\~no and G. Moore, ``The Donaldson-Witten 
function for gauge groups
of rank larger than one," hep-th/9802185, Commun. Math. Phys. {\bf 199}
(1998) 25.

\bibitem{mmtwo} M. Mari\~no and G. Moore, 
``Donaldson invariants for non-simply connected
manifolds," hep-th/9804114, Commun. Math. Phys. {\bf 203} (1999) 249.

\bibitem{mmthree} M. Mari\~no and G. Moore, ``Three-manifold 
topology and the Donaldson-Witten partition function,'' 
Hep-th/9811214, Nucl. Phys. {\bf B 547} (1999) 569.

\bibitem{mmp} M. Mari\~no, G. Moore and G. Peradze, 
``Superconformal invariance
and the
geography of four-manifolds," hep-th/9812055, Commun. Math. Physics. 
{\bf 205} (1999) 691; ``Four-manifold geography and
superconformal
symmetry," math.DG/9812042, Math. Res. Lett. {\bf 6} (1999) 429.

\bibitem{mt} G. Meng and C.H. Taubes, ``${\underline {SW}}$=Milnor torsion,'' 
Math. Res. Lett. {\bf 3} (1996) 661.

\bibitem{mw} G. Moore and E. Witten, ``Integration over
the $u$-plane in Donaldson theory," hep-th/9709193, Adv. Theor. Math. Phys.
{\bf 1} (1997) 298.

\bibitem{vm} V. Mu\~noz, ``Gromov-Witten invariants of the moduli of 
bundles on a surface,'' math.AG/9910105. ``Ring structure of the Floer 
cohomology of $\Sigma\times {\bf S}^1$, dg-ga/9710029, Topology {\bf 38}
(1999) 517.

\bibitem{rw} L. Rozansky and E. Witten, ``HyperK\"ahler geometry and 
invariants of three-manifolds,'' hept-h/9612216, Selecta. Math. (N.S.) 
{\bf 3} (1997) 401.  

\bibitem{sw} N. Seiberg and E. Witten,
``Electric-magnetic duality, monopole condensation, and confinement in
$N=2$ supersymmetric Yang-Mills
theory,''
hep-th/9407087, Nucl. Phys. {\bf B 426} (1994) 19. 
``Monopoles, duality and chiral symmetry breaking in $N=2$
supersymmetric QCD,''
hep-th/9408099, Nucl. Phys. {\bf B 431} (1994) 484.

\bibitem{th} M. Thaddeus, 
``Conformal field theory and the cohomology of the
moduli space of stable bundles," J. Diff. Geom. {\bf 35} (1992) 131.

\bibitem{tqft} E. Witten,
``Topological Quantum Field Theory,''
Commun. Math. Phys. {\bf 117} (1988)
353.

\bibitem{mfm} E. Witten, ``Monopoles and
four-manifolds,''  hep-th/9411102; Math. Res. Letters {\bf 1} (1994) 769.



\end{thebibliography}
\end{document}